# Polaron-assisted dielectric relaxation processes in donor-doped BaTiO$_3$-based ceramics

T.H.T. Rosa[1], M.A. Oliveira[1,2], Y. Mendez-González[3], F. Guerrero[4], R. Guo[5], A.S. Bhalla[5], J.D.S. Guerra[1,*]

[1]Grupo de Ferroelétricos e Materiais Multifuncionais, Instituto de Física, Universidade Federal de Uberlândia, Uberlândia – MG, 38408-100, Brazil

[2]Departamento de Física, Universidade Estadual Universidade do Estado de Minas Gerais, Passos (UEMG), Passos – MG, 37900-106, Brazil

[3]Instituto de Fusión Nuclear "Guillermo Velarde", Universidad Politécnica de Madrid, José Gutiérrez Abascal 2, Madrid E-28006, Spain

[4]Departamento de Física de Materiais, Instituto de Ciências Exatas, Universidade Federal do Amazonas, Manaus – AM, 69077-000, Brazil

[5]Multifunctional Electronic Materials and Devices Research Lab., Department of Electrical and Computer Engineering, College of Engineering, The University of Texas at San Antonio, San Antonio 78249, TX, USA.

Perovskite structure materials based on the Ba$_{1-x}$Gd$_x$TiO$_3$ system, where $x$ = 0.001, 0.002, 0.003, 0.004 and 0.005, were prepared *via* the Pechini's chemical synthesis route. The dielectric properties have been analyzed over a wide temperature and frequency range, revealing a significant contribution of the conduction mechanisms in the dielectric response of the studied ceramics. In fact, by using the Davidson-Cole formalism, the observed electrical behavior was found to be associated with relaxation processes related to intrinsic defects mobility promoted by a thermally-activated polaronic mechanism. The obtained values of the activation energy for the relaxation processes, estimated from the Arrhenius' law for the mean relaxation time, revealed a decrease from 0.29 up to 0.21 eV as the Gd-doping concentration increases, which suggests the conduction process to be associated with the polaronic effects due to the coexistence of Ti$^{4+}$ and Ti$^{3+}$ ions in the structure. Analysis from the conductivity formalism, by using the Jonscher's universal power-law, confirmed the polaron-type conduction mechanism for the dielectric dispersion, as suggested by the dielectric analysis, being the nature of the hopping mechanism governed by small polaron hopping (SPH) charge transport in the studied Ba$_{1-x}$Gd$_x$TiO$_3$ ceramics.

**Keywords:** Dielectric relaxation, BaTiO$_3$ ceramics, Conduction mechanisms, Polaron

---

* Corresponding author. *E-mail address*: santos@ufu.br (J.D.S. Guerra).

# 1. Introduction

There has been an increased interest on lead-free materials, in both ceramic and thin films, because of their potential application as active elements in many areas of solid-state physics such as electro-electronic devices and photocatalysts [1,2]. This is the case of $BaTiO_3$-based materials showing not only an excellent ferroelectric, piezoelectric and dielectric response but also exceptional semiconductor properties [3,4], including the manifestation of positive temperature coefficient of resistance (PTCR) effect promoted by the incorporation of donor dopants [5,6]. These unique features have strengthened the applications of the barium titanate in lead-free devices such as thermistors, piezoelectric and electro-optical devices as well as multilayer ceramic capacitors (MLCs) [3,4,7].

The modified $BaTiO_3$ perovskite has been investigated to technologically advance high-quality lead-free materials, and its properties can be tailored by substituting or doping host cations at the *A* and/ or *B*-site of the structure [2,7–12]. For instance, semiconductor characteristics can be induced by the incorporation of $La^{3+}$, $Er^{3+}$ or $Gd^{3+}$ ions at the Ba-site, as well as the substitution of $Ti^{4+}$ cation by $Nb^{5+}$ or $Ta^{4+}$ elements [7–10]. Additionally, the formation of cation vacancies and oxygen defects from the introduced dopants might change the electron-hole recombination rate through trap sites complementation, which is one of the factors through which the physical properties (i.e., photocatalytic and semiconductor activities) could be improved [11]. Nevertheless, less effort has been indeed devoted to the understanding of the influence of the conduction mechanisms on the dielectric characteristics in BT-based compounds. Such advances could be crucial for getting further insights into the optimal physical properties that this material finally shows after doping. In particular, little has been found in the literature regarding results that explore semiconducting properties taking into account doping at the *B*-site.

The possibility of the conduction could be due to the hopping process mainly related to oxidation/reduction processes, which are characteristic of some constituent elements in the composition as well as extrinsic conductivity associated with charge transfer mechanisms, oxygen vacancies and intrinsic ionic conduction at high temperatures [13]. For perovskite materials, the polaron relaxation exists generally at low temperatures in crystals with structural defects [14]. According to the literature reports [13–17], activation energy values between 0.03–0.52 eV are associated with thermally-activated polaron mechanisms promoted by the electron and/or hole-phonon interaction. This phenomenon, that has been associated with the interaction between the electron and surrounding lattice vibrations, can originate due to the interaction of hopping mechanisms of electrons (or holes) that interact with the vibration of the crystal lattice and this interaction behaves as a "particle" (polaron) that can move with excitation from certain energy values, and hence propagate along the structure [18].

From the dielectric properties point of view, several models have been reported in the literature to account for the relaxation processes in perovskite-type ferroelectrics [19], where the dielectric response is governed by a strong dependence of the dielectric permittivity with frequency. The influence of different point defects (e.g., cation and oxygen vacancies) that appear in the material during the sintering process, has been widely accepted as one of the main causes for the observed behavior. In fact, the coexistence of different relaxation mechanisms promotes different activation energy ($E_a$) barriers, which can be individually related to the involved process. From the fundamental point of view, $E_a$ values laying in the 0.79–1.2 eV interval are commonly attributed to the conduction/relaxation behaviors coming from the grain and grain-to-grain boundary regions [20]. Activation energy values close to 0.79 eV infer that the intrinsic electronic conduction, in the lower temperature region, are related to the charge transport mechanism promoted by the presence of vacancy defects, which according to M. Li *et al.* could be generated by the volatilization of some chemical elements during sintering of the ceramic samples [21]. In the high temperature region, however, several authors have reported that activation energy values around 1.2 eV are associated with the conduction process from cationic as well as doubly ionized oxygen vacancies [22–26]. On the other hand, Stumpe *et al.* [27] considered the dielectric relaxation process with extremely high dielectric permittivity as the combined effect from both bulk and surface properties in both $SrTiO_3$ and $BaTiO_3$ based perovskites, namely the Maxwell-Wagner polarization effect. Most recently, electrical conductivity studies carried out in ZnO-based nanocomposites revealed that the electrical conduction mechanism is also related to a charge-hopping process, confirming the influence of the grain and grain-boundary contributions in the electrical conduction processes [28]. However, it can be seen that the real nature of the observed strong dielectric dispersion in perovskite-type materials remains up today under exhaustive discussion and, therefore, has to be still carefully investigated in detail. This effect has been ascribed to additional conduction mechanisms, which directly affect the real polarization response.

In this work, the polarization relaxation processes and their involved conduction mechanisms have been investigated in Gd-modified $BaTiO_3$ ceramics ($Ba_{1-x}Gd_xTiO_3$) from the dielectric response in a wide range of temperature and frequency. The conductive effects provided by the doping with donor ions at the *A*-site of the perovskite structure, whose study has not yet been reported in the literature, were carefully investigated to better understand the factors that influence and determine the semiconductor properties of such materials.

2. **Experimental procedure**

$Ba_{1-x}Gd_xTiO_3$ ceramic samples (where $x$ = 0.001, 0.002, 0.003, 0.004 and 0.005) were obtained *via* the Pechini's chemical synthesis method [29], as reported elsewhere [30], using high-purity chemical grades ($C_6H_8O_7$, Synth: 99.5 %; $HOCH_2CH_2OH$, Synth: 99.0 %; $Ti[OCH(CH_3)_2]_4$, Sigma-Aldrich, 99.9 %; $Gd_2O_3$, Alfa Aesar: 99.9 % and $Ba(CH_3COO)_2$, Synth: 99.0 %). A homogeneous solution of titanium citrate was firstly produced by mixing the citric acid ($C_6H_8O_7$) and titanium isopropoxide (($Ti[OCH(CH_3)_2]_4$) at constant stirring and 60 $^o$C. After dissolving the gadolinium oxide ($Gd_2O_3$) and barium acetate ($Ba(CH_3COO)_2$ separately in nitric acid ($HNO_3$, 65.0%, Quimis) and citric acid solutions, respectively, both the gadolinium oxide and titanium citrate were added to the barium acetate and mixed at 60 $^o$C, until the formation of a homogenous solution. The polymerization process was promoted by dissolving ethylene glycol ($HOCH_2CH_2OH$) in the previously formed solution, followed by a rigorous control of the pH through the addition of ammonia hydroxide ($NH_3OH$). After several thermal treatments, a stable powdered polymeric resin was obtained and calcined at 700 $^o$C for 2.5 h. The obtained powders were then uniaxially and isostatically pressed under 2.0 MPa and 200 MPa, respectively, forming disc-shaped "green" pellets. Finally, the samples were sintered at 1300 $^o$C for 2 h. Structural analyses, carried out from X-ray diffraction (XRD) measurements, as previously reported [30], confirmed the formation of the perovskite structure with tetragonal symmetry (P4*mm* space-group) for all the analyzed compositions, and the results revealed a strong influence of the lanthanum concentration on the lattice parameters as well as the unit-cell distortion (tetragonality). For the electrical measurements, silver-painted electrodes were applied on the parallel opposite surfaces of the samples by a heat treatment at 590 $^o$C. Dielectric measurements were performed by using a HP4194A Impedance Analyzer in a wide frequency (100 Hz–1 MHz) and temperature (25–400 $^o$C) range. The samples were hereafter labeled as BGT001, BGT002, BGT003, BGT004 and BGT005 for $x$ = 0.001, 0.002, 0.003, 0.004 and 0.005, respectively.

3. **Results and discussion**

*3.1. Dielectric response*

The dielectric properties have been analyzed from the temperature dependence of the real and imaginary components of the dielectric permittivity ($\varepsilon'$ and $\varepsilon''$, respectively) and results are shown in Fig. 1 for all the studied compositions, in a wide frequency range.

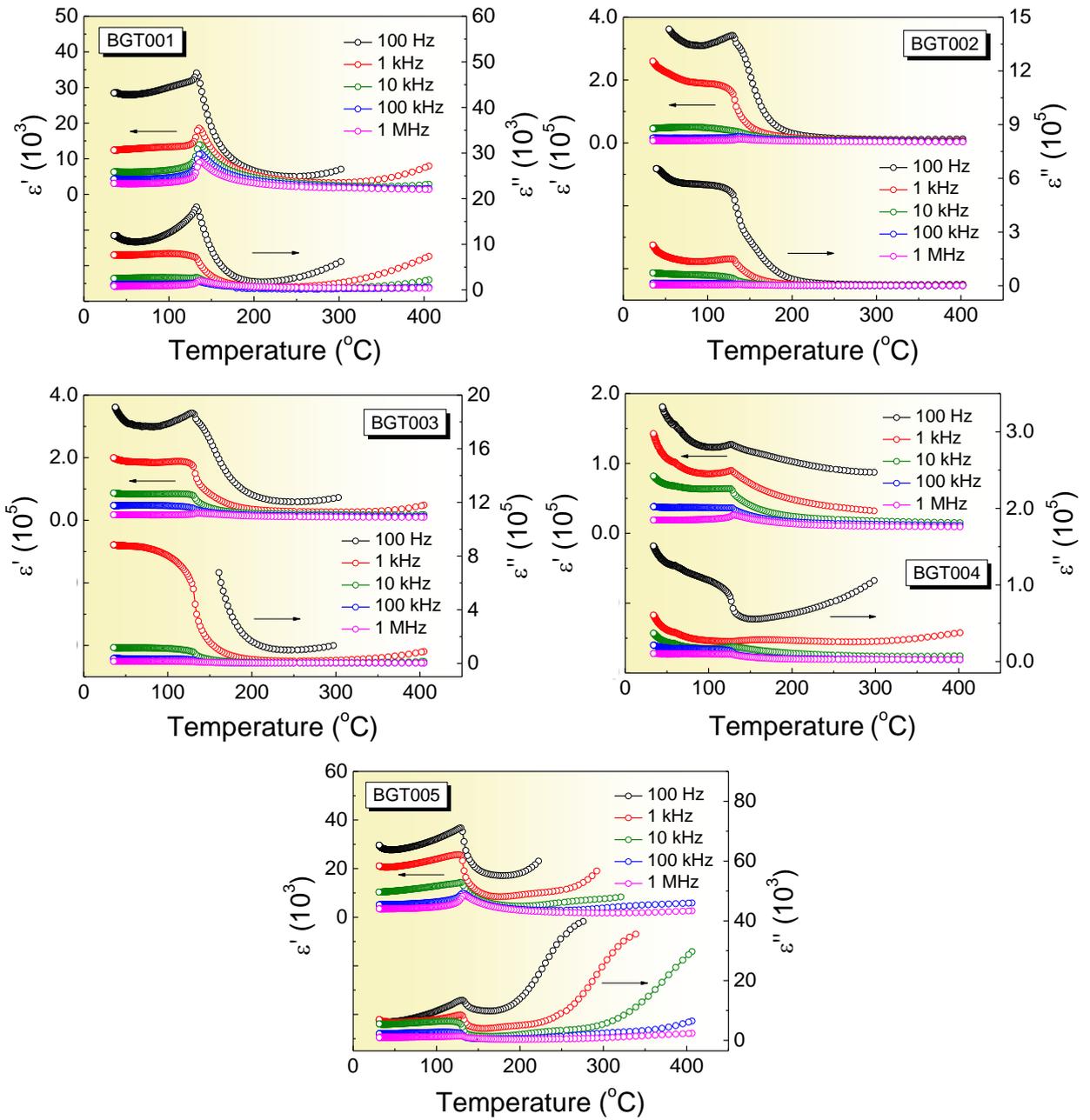

Figure 1. Temperature dependence of the real ($\varepsilon'$) and imaginary ($\varepsilon''$) components of the dielectric permittivity for the studied BGT00$x$ compositions, at different selected frequencies.

As can be seen, typical paraelectric-ferroelectric phase transitions can be identified for all the cases, from the maximum dielectric permittivity ($\varepsilon_m$) observed around the Curie temperature ($T_C$). Fig. 2 shows the composition dependence of both $\varepsilon_m$ and $T_C$ obtained for the studied compositions, at 1 kHz, which have been extracted from Fig. 1. While $\varepsilon_m$ depicts its maximum value for the BGT003 composition, $T_C$ reveals a decreasing behavior with the increase of the $Gd^{3+}$ content, which confirms the high solubility of the doping cation in the perovskite structure, by substituting the $Ba^{2+}$ ion at the $A$-site [31]. Such behavior observed in $T_C$ has been also reported for other $BaTiO_3$, $SrTiO_3$ and $PbTiO_3$-based ceramics [32,33], and the results have been explained by a direct correlation between the electrical properties and the microstructural characteristics of the sample [33]. Indeed, according to Keizer *et al.* [33], a decrease in $T_C$ has been observed in connection with a decrease in the average grain-size induced by internal compressional stresses, when the sample is subjected to an applied external hydrostatic pressure. In this context, a decrease in the grain-size promotes an increase in the compressional stress in specific directions in the material, and this behavior has been used in analogy to the fact that the effect which promotes the decrease in the grain-size is similar to the increase in the hydrostatic pressure. Therefore, the observed decrease in $T_C$ for the studied compositions could be a consequence of the decrease in the average grain-size from BGT001 to BGT005, as reported elsewhere [30].

Furthermore, considering that the reported theoretical value for the Curie temperature in the pure $BaTiO_3$ system and some rare-earth modified BT compositions are found to be between 120 °C and 135 °C, depending on the used synthesis method for the samples' preparation [34,35], it can be observed that the obtained values for $T_C$ in the studied compositions are in agreement with the reported results in the literature [36,37].

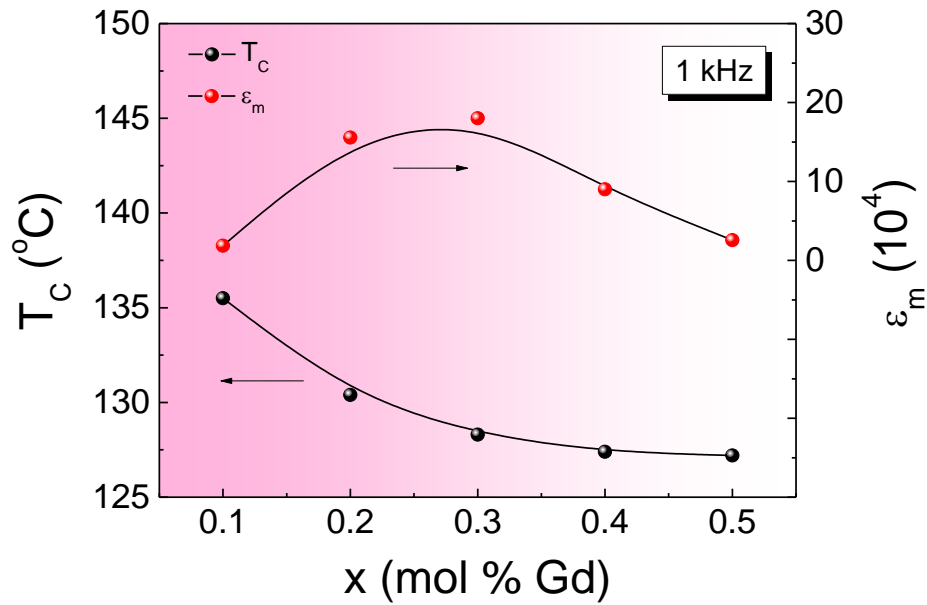

Figure 2. Composition (*x*) dependence of the maximum dielectric permittivity ($\varepsilon_m$) and the Curie temperature ($T_C$) for the studied BGT00*x* compositions at 1 kHz.

Back to the temperature dependence of the dielectric permittivity (Fig. 1), it can be seen that the dielectric response is governed by a strong frequency dispersion, which is shown to be more prominent for temperatures below $T_C$ for all the analyzed compositions. This behavior, which is indicative of the presence of a highly conductive component in the studied material, directly affects the expected behavior of crucial macroscopic physical parameters such as polarization and dielectric permittivity [38]. In this context, due to the high conductivity observed at room temperature, it is difficult to determine the true ferroelectric hysteresis loop in the samples because of the conductive effects induced by structural defects, including vacancies, chemical impurities, etc. As can be seen, the maximum dielectric permittivity ($\varepsilon'_m$) increases with increasing the Gd content, up to the BGT003 composition, and then decreases for higher concentrations. It can be observed that all the analyzed compositions presented dielectric permittivity values too much higher than those reported for the pure BT system, which has been ascribed to the prominent conduction character of the lanthanide (Ln) elements when inserted into the BT perovskite structure [39]. On the other hand, it can be noted that this conductive behavior reveals to be stronger in the imaginary dielectric permittivity ($\varepsilon''$), which is directly related to the dielectric losses of the system. In fact, it can be observed in Fig. 1 that, for some compositions, the conductive nature of the system does not even allow to define the characteristic peak of $\varepsilon''$, at the maximum imaginary dielectric permittivity at $T_C$ ($\varepsilon''_m$).

It is well known from the current literature that when doping the $BaTiO_3$ system with moderate amounts of donor impurities (i.e., $0.1 \leq x \leq 0.3$ mol % Ln) [40], the increase in conductivity at low temperatures ($T \ll T_C$) is attributed to the excess of conduction electrons created due to the charge unbalance promoted from the incorporation of donor ions into the crystal lattice [41]. However, for dopant concentrations above 0.3 mol %, the electrical conductivity behavior remains not well understood [42]. Some authors, indeed, have reported that the inclusion of lanthanide ions in the $BaTiO_3$ crystal lattice for concentrations higher than $x=0.003$ (0.3 mol %) promotes a decrease in the dielectric permittivity due to the occurrence of chemical defects [43]. These defects are attributed to the excess of cation vacancies, which promote a decrease in the electron mobility for the conduction band [44], thus causing an increase in the resistivity of the material and thus losing the semiconductor properties. In this context, in order to better understand the nature of the observed conductive effects, which strongly affected the dielectric response of the studied compositions, the frequency dependence of the dielectric response has been also analyzed, the approach to which will be presented as follows.

From the obtained results for the temperature dependence of the dielectric permittivity, shown in Fig. 1 at several frequencies, the frequency dielectric dispersion has been obtained and the results are depicted in Fig. 3, where the frequency dependence of the real ($\varepsilon'$) and imaginary ($\varepsilon''$) dielectric permittivity obtained at 60 °C is shown for all the compositions, as an example of the obtained results for all the analyzed temperatures, below the ferroelectric-paraelectric phase transition temperature ($T_C$).

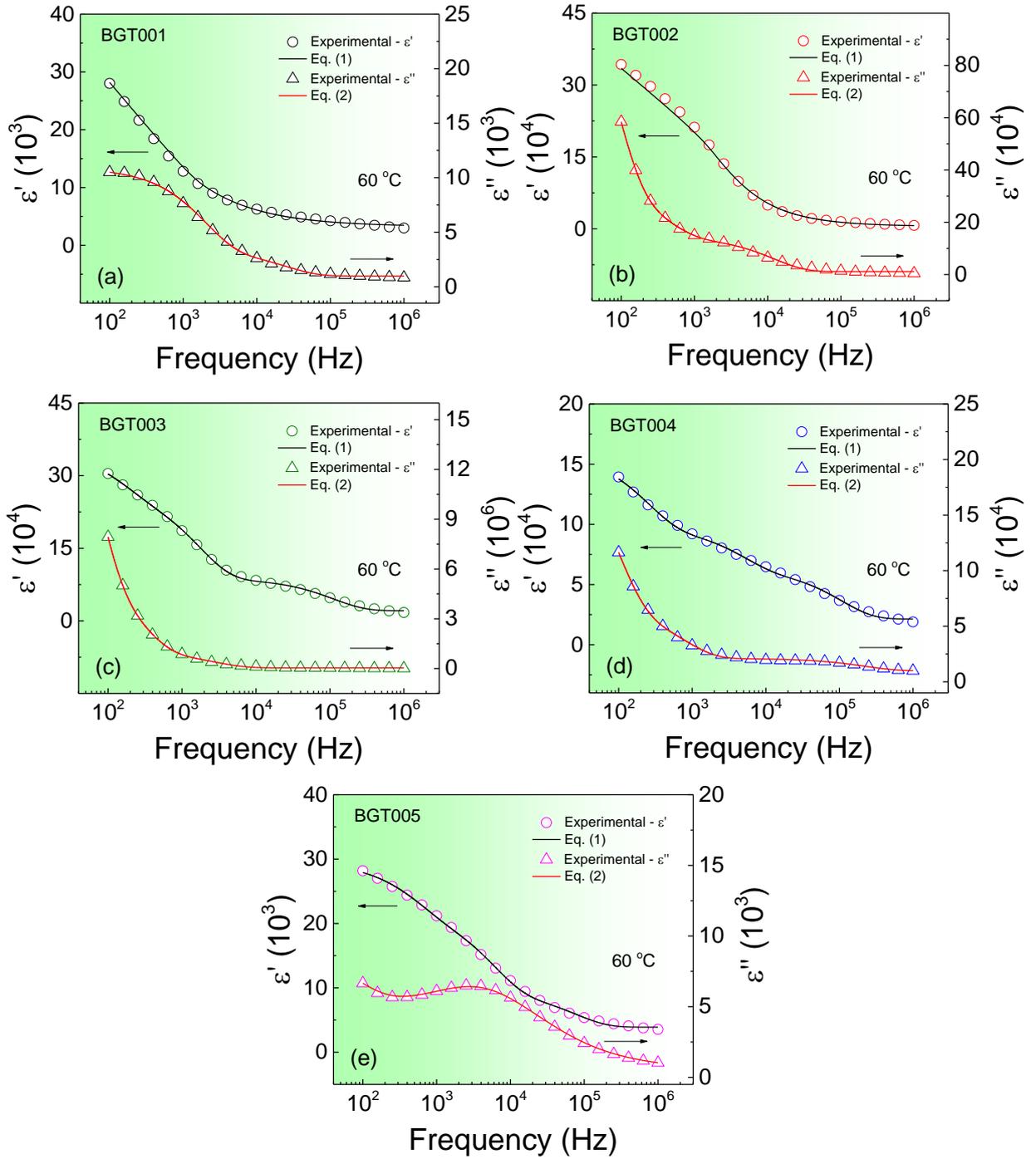

Figure 3. Frequency dependence of the real (ε') and imaginary (ε'') components of the dielectric permittivity for the studied compositions at 60 °C.

To account for such observed dispersion, several models based on different relaxation distribution functions have been proposed [45–50], and have been successfully accepted for the description of the relaxation processes behind the frequency dispersion in the dielectric permittivity in perovskite structure systems [51,52]. In this work, the best fitting of the experimental data has been obtained from the Davidson-Cole distribution function [47], for which the frequency dependence of the complex dielectric permittivity ($\varepsilon^* = \varepsilon' + j\varepsilon''$) can be described by the Eqs. (1) and (2) for ε' and ε'', respectively. The $\varepsilon_s$ and $\varepsilon_\infty$ parameters are, respectively, the low (static) and high-frequency dielectric permittivities, $\phi = arctan(\omega\tau)$ (where $\omega = 2\pi f$, being $f$ the measurement frequency and $\tau$ represents the mean relaxation time for the relaxation process) and $\beta$ is a constant characteristic of the material, having a value in the $0 \leq \beta \leq 1$ interval. When $\beta = 1$, the Davidson-Cole distribution function yields the classical Debye's relaxation expression [45].

$$\varepsilon' = \varepsilon_\infty + (\varepsilon_s - \varepsilon_\infty)(\cos\phi)^\beta \cos(\phi\beta) \qquad (1)$$

$$\varepsilon'' = (\varepsilon_s - \varepsilon_\infty)(\cos\phi)^\beta \sin(\phi\beta) \qquad (2)$$

Formally, the Davidson-Cole's model represents a generalization of the Cole-Cole's formalism, by introducing an additional fractional exponent, and considers a more complex distribution function for the relaxation time, thus offering a better description in the frequency domain than the expected from the classical Debye and Cole-Cole relaxation equations [53]. The Eqs. (1) and (2) have been then used to analyze the frequency spectra of the dielectric permittivity and the results are also shown in Fig. 3, represented by solid lines, from which the characteristic relaxation time have been estimated for each analyzed temperature.

For dielectric relaxation processes, the relaxation frequency is given because the polarization arises from the same thermally activated processes that give rise to the DC conductivity. Therefore, the relaxation time will depend on temperature by an exponential factor following the so-called Arrhenius' law, according to the Eq. (3), where $T$ represents the absolute temperature, $\tau_0$ is the pre-exponential factor, $k_B$ is the Boltzmann constant and $E_a$ is the activation energy for the relaxation process.

$$\tau = \tau_0 \exp\left(\frac{E_a}{k_B T}\right) \qquad (3)$$

Fig. 4 shows the converse temperature dependence of the characteristic relaxation time ($\ln\tau$), obtained from the fitting of the experimental data with the Davidson-Cole relation, expressed by the Eqs. (1) and (2). Solid-lines in Fig. 4 represent the fitting results with the Eq. (3) and confirm the expected linear behavior from the Arrhenius's law. The obtained values for the activation energy ($E_a$) for all the studied compositions are reported in Table 1 and reveal to be in the 0.21–0.29 eV range for the analyzed temperature interval.

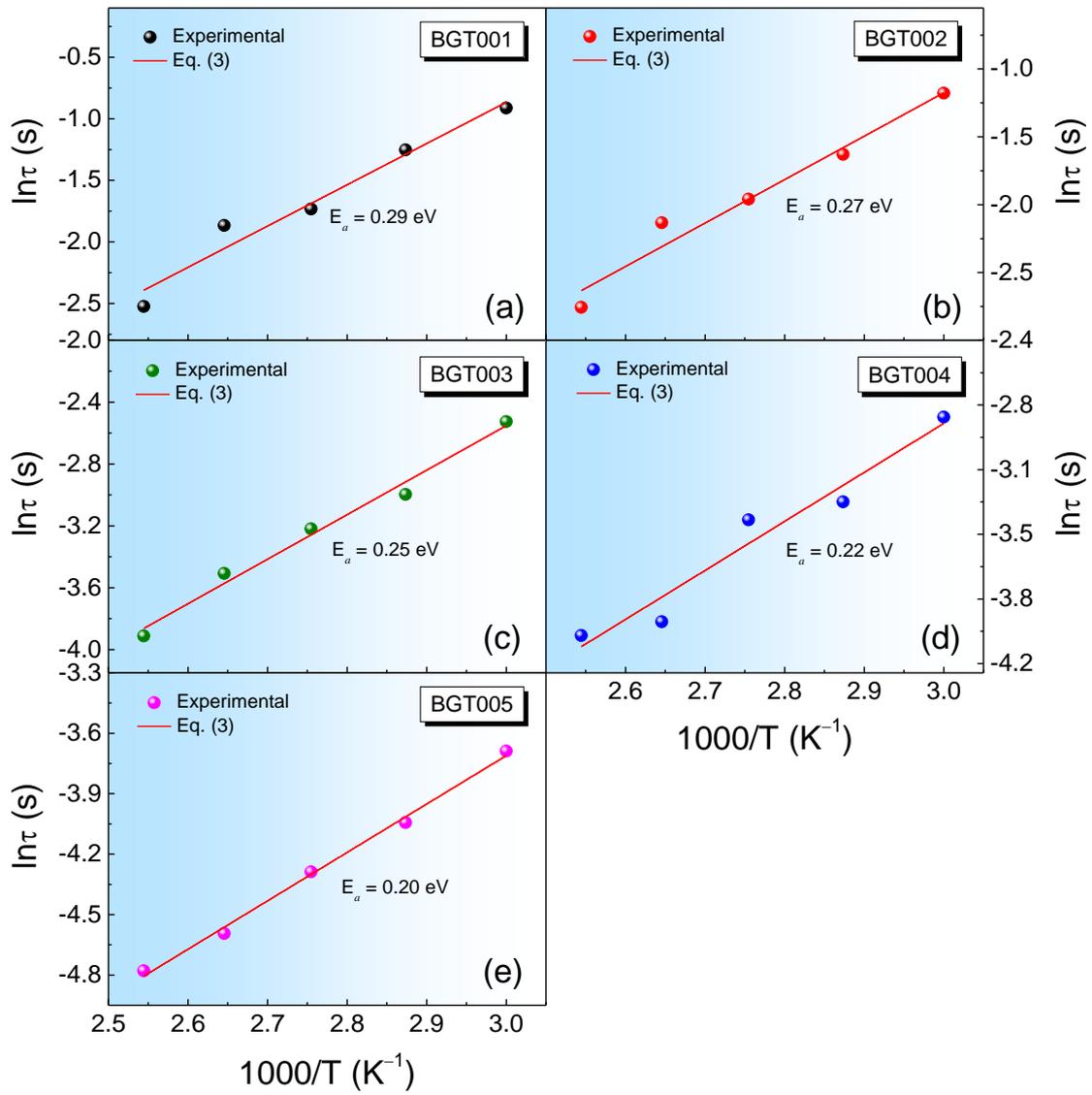

Figure 4. Temperature dependence of the characteristic relaxation time (lnτ), obtained from the fitting of the experimental data in Fig. 3 with the Davidson-Cole distribution function. Solid lines represent the fitting with the Arrhenius' law, according to the Eq. (3).

Table 1. Activation energy values obtained from both relaxation ($E_a$) and conductivity processes ($E_{DC}$). The $E_0$ parameter refers energy necessary for the polaron release.

| Composition | $E_a$ (eV) | $E_{DC}$ (eV) | $E_0$ (eV) |
|---|---|---|---|
| **BGT001** | 0.28(8) | 0.31(2) | 0.046(7) |
| **BGT002** | 0.27(5) | 0.29(6) | 0.040(6) |
| **BGT003** | 0.24(8) | 0.26(7) | 0.037(9) |
| **BGT004** | 0.22(8) | 0.24(3) | 0.029(8) |
| **BGT005** | 0.20(7) | 0.22(0) | 0.026(4) |

According to literature reports, activation energy values between 0.03 eV and 0.52 eV may be associated with the hopping mechanisms of charge carriers, mainly related to oxidation/reduction processes, which are characteristics of some chemical elements in the composition [13–17], as well as the formation of polarons promoted by the electron-phonon (or hole-phonon) interaction intensified by structural deformations [54]. Activation energy values associated with thermally activated polaron mechanisms have been also reported for the $BaTiO_3$ system (thin films and single crystals) [55,56]. In this way, the obtained values for the activation energy of the studied samples in the present work could be associated with conduction processes characteristic of thermally activated polarons. This phenomenon, related to the electron-phonon interaction, may be caused by the interaction of hopping mechanisms of electrons (or holes) that interact with the crystal lattice vibrations and this interaction behaves like a "particle" (polaron), which can move, and propagate throughout the structure, under external excitation from certain energy values [18].

From a fundamental point of view, a polaron is considered to be a quasiparticle, which describes the interaction between electrons and ions in a solid material resulting in a bound state [57]. When the electron moves slowly within the crystal, it can produce a local deformation in the surrounding lattice because of the interaction with the closest ions. Such deformation moves with the electron through the crystal, thus creating a polaron that propagates in a specific direction [57,58]. Since in high conductivity materials the electronic mobility is strongly affected by the creation of polarons, they become extremely relevant for practical applications where the charge transport reveals to be an essential mechanism for the design of electronic devices. Fig. 5 depicts a schematic representation of a polaron, showing the polarization induced by the localization of an electron at a lattice site, deforming the surrounding lattice.

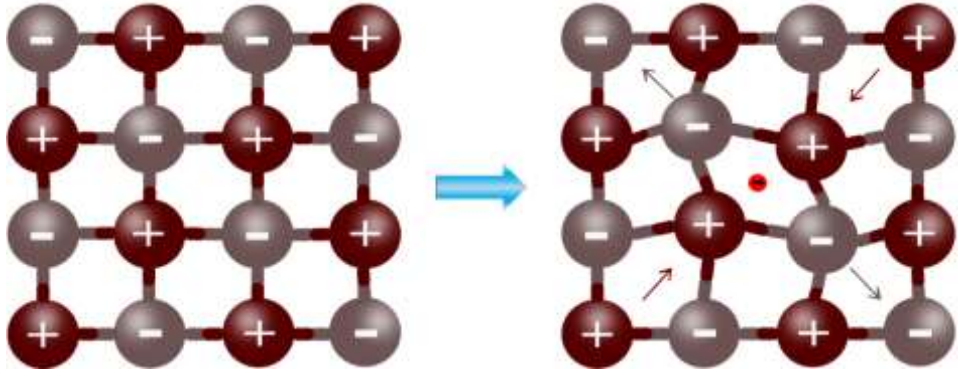

Figure 5. Schematic diagram of a polaron, showing a bi-dimensional crystal structure before (left) and after (right) the influence of a moving electron deforming the surrounding lattice.

Polarons originate mainly in ionic crystals, as a result of the extremely high Coulomb interaction between electrons and ions in the crystalline lattice [58,59]. The ions move out from their equilibrium positions in order to successfully screen the electron's charge, and this mechanism is commonly identified as the phonon's cloud. In this case, the deformation produces a local polarization, which gives rise to the so-called polaron [59]. Therefore, a polaron can be seen as an electron interacting with a cloud of phonons that propagates, producing a conduction mechanism in this type of material. As a consequence, the effective mass of the electron appears to increase, thus decreasing its mobility [60–62]. Polarons can be classified according to their effective interaction as 'large polarons' and 'small polarons' [63,64]. Large polarons are characterized by a weak interaction between electrons and the lattice, so that the increase in effective mass is not very significant and, therefore, they can easily move through the crystal lattice [63]. In other words, the radius of a polaron is much larger than the lattice parameter of the material. Small polarons, on the other hand, are characterized by a strong electron-phonon interaction and the electron is tightly bound to a lattice ion, since it does not have energy enough to overcome the potential barrier separating adjacent ions [64]. In this case, the displacement of the polaron from one position to another in the lattice occurs either by thermal activation, at high temperatures, or by tunneling, at lower temperatures. This type of polaron is characteristic in dielectric materials where the polaron's displacement can be, therefore, in the order of the unit-cell lattice parameter and obeys the Arrhenius' law behavior for the hopping mechanism, whose exponent becomes strongly dependent on the binding energy and temperature [18,58,59].

From the experimental point of view, the polaron mechanism can be identified from the low-frequency dielectric dispersion and can be explained by considering the charge neutrality of the unit-cell, according to the scenario shown below. The charge neutrality of the unit-cell in the perovskite structure (which is our case) can occur through the following three mechanisms: $Ba^{2+}=Gd^{3+}+e^{-}$, $Ti^{4+}+e^{-}=Ti^{3+}$ and $O^{2-}=½O_2+V_O+2e^{-}$. Noting that, in this work, the BT system is being doped with donor ions at the *A*-site of the perovskite structure, the substitution of the $Ba^{2+}$ by the $Gd^{3+}$ ion promotes a charge unbalance in the unit-cell lattice. Therefore, in order to achieve the charge neutrality, as previously pointed out, the possible mechanisms to be considered are the following: *i*- the creation of positive vacancies at the *A*- and/or *B*-site or *ii*- the reduction of the titanium ion from $Ti^{4+}$ to $Ti^{3+}$. Although both mechanisms are favorable, since the $Ti^{4+}$ is a transition metal cation, the energy required to be reduced is much lower than the one required for the creation of positive vacancies, as previously reported [9]. According to the literature, Ba and Ti vacancies are possible to be created only for $E_a$ values above 1.30 eV [9,65,66], being these values much higher than those obtained in this work (0.21–0.29 eV). In this context, the electron hopping mechanism from the $Ti^{4+}$ ion interacts with the lattice phonons, thus promoting the formation of polarons, which propagate through the crystal lattice. Hence, the observed behavior in the dielectric properties obtained for the studied system seems to be associated with the polaronic effects due to the coexistence of $Ti^{4+}$ and $Ti^{3+}$ ions in the structure. This result is in agreement with previously reported work by Iguchi *et al.* for lanthanum-modified $BaTiO_3$ ceramics and other perovskite-based systems [15]. The activation energy values obtained in the present work confirm, therefore, the polaron conduction mechanism for the analyzed BGT00*x* compositions, due to the presence of a polaronic deformation caused by the creation of $Ti^{3+}$ ions, which distort the local crystal lattice as the polaron displaces along the $Ti^{4+}$–O–$Ti^{3+}$ bonds.

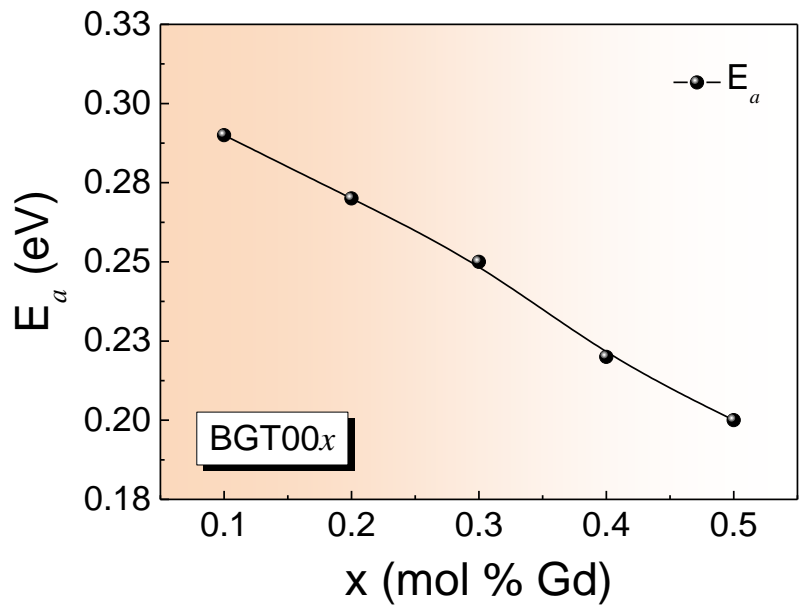

Figure 6. Composition (*x*) dependence of the activation energy ($E_a$) for the studied BGT00*x* compositions.

Fig. 6 shows the composition (*x*) dependence of the activation energy ($E_a$) for the studied samples. It is observed that $E_a$ decreases with the increase of the Gd content, which can be explained by the charge mismatch introduced by the donor ion. The increase in the doping concentration promotes an excess of the positive charges, which leads to an increase in the conduction mechanism favored by the $Ti^{4+}$–$Ti^{3+}$ exchange mechanism, thus promoting a decrease in the activation energy of the system. Therefore, the increase in the Gd content seems to act in favor of the electron-phonon coupling.

*3.2. Conductivity analysis*

In order to better clarify the real contribution of the observed dielectric relaxations and, therefore, the real nature of the above discussed conduction mechanism, the conductivity formalism has been now explored. The frequency response for the real component of the complex conductivity (σ) was then analyzed, from the Jonscher's universal power-law expressed by the Eq. 4 [50], where $\omega$ is the frequency (being $\omega = 2\pi f$), $\sigma_{DC}$ is the DC conductivity, $A$ is a temperature-dependent pre-exponential factor and *s* the frequency exponent ($0 < s < 1$), which is commonly associated with the slope of the conductivity curve and measures the interaction between mobile charges. The AC conduction mechanism can be, therefore, determined from the frequency response of σ at different temperatures.

$$\sigma(\omega) = \sigma_{DC} + A\omega^s \tag{4}$$

Fig. 7 shows the frequency dependence of σ for the studied compositions at 60 °C, as an example of the obtained results for all the analyzed temperatures, below the ferroelectric-paraelectric phase transition temperature ($T_C$). From the fitting of the experimental data with the Eq. 4, the characteristics parameters $\sigma_{DC}$ and *s* have been extracted for all the studied compositions. Several models have been proposed in the literature in order to investigate the conduction mechanisms in highly frequency dispersive perovskite-based materials [67–70], where the variation of the frequency exponent (*s*) with temperature plays the main role. In fact, classical charge carriers hopping characterized by a continuous increase in *s* with the increase in temperature has been related to small polaron hopping (SPH) transport [67,71,72], which is common in strong electron-phonon coupling systems [63]. On the other hand, the decrease in *s* with temperature has been associated with a correlated barrier hopping (CBH) mechanism [68,73,74]. This model, proposed by Elliot [68], considers the hopping of two electrons over a potential barrier between defect sites, being the barrier's height correlated with the intersite separation. When a minimum is observed in the temperature dependence of *s*, the conduction can be related to a large polaronic hopping (LPH) charge transport [69,75,76], where the overlapping of the polaron wells of two sites is observed, thus reducing the hopping energy. Furthermore, the conduction mechanism has been proposed to be governed by quantum mechanical tunneling (QMT) processes when the frequency exponent is temperature-independent [70,72,77]. According to this model, since the electron does not have energy enough to overcome the potential energy barrier (from the classical mechanics point of view), it passes across the potential barrier with a lower energy [70]. Therefore, the temperature dependence of the frequency exponent ($s$), has been used to understand the nature of predominant conduction mechanism of the studied system in the present work.

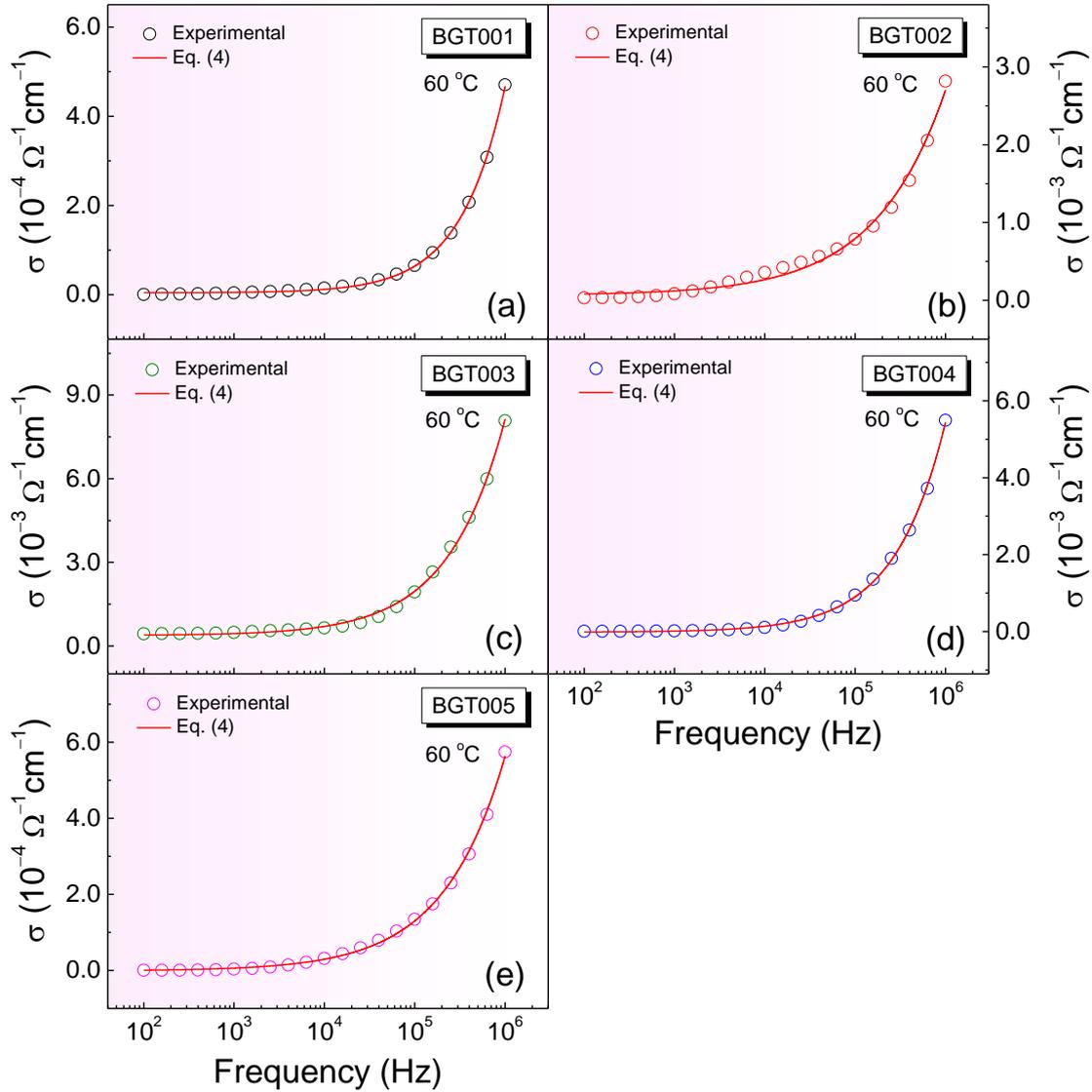

Figure 7. Frequency dependence of the real component of the electrical conductivity for the studied compositions at 60 °C.

Fig. 8 depicts the temperature dependence of the frequency exponent ($s$) for all the studied compositions. Results reveal a direct variation in $s$ with temperature for all the samples, and indicate that the nature of the hopping mechanism is governed by small polaron hopping (SPH) charge transport in the studied $Ba_{1-x}Gd_xTiO_3$ ceramics. Therefore, it can be seen that the obtained results from the conductivity formalism confirm the previously suggested polaron-type conduction mechanism by the dielectric analysis, being associated to small polaron hopping charge transport.

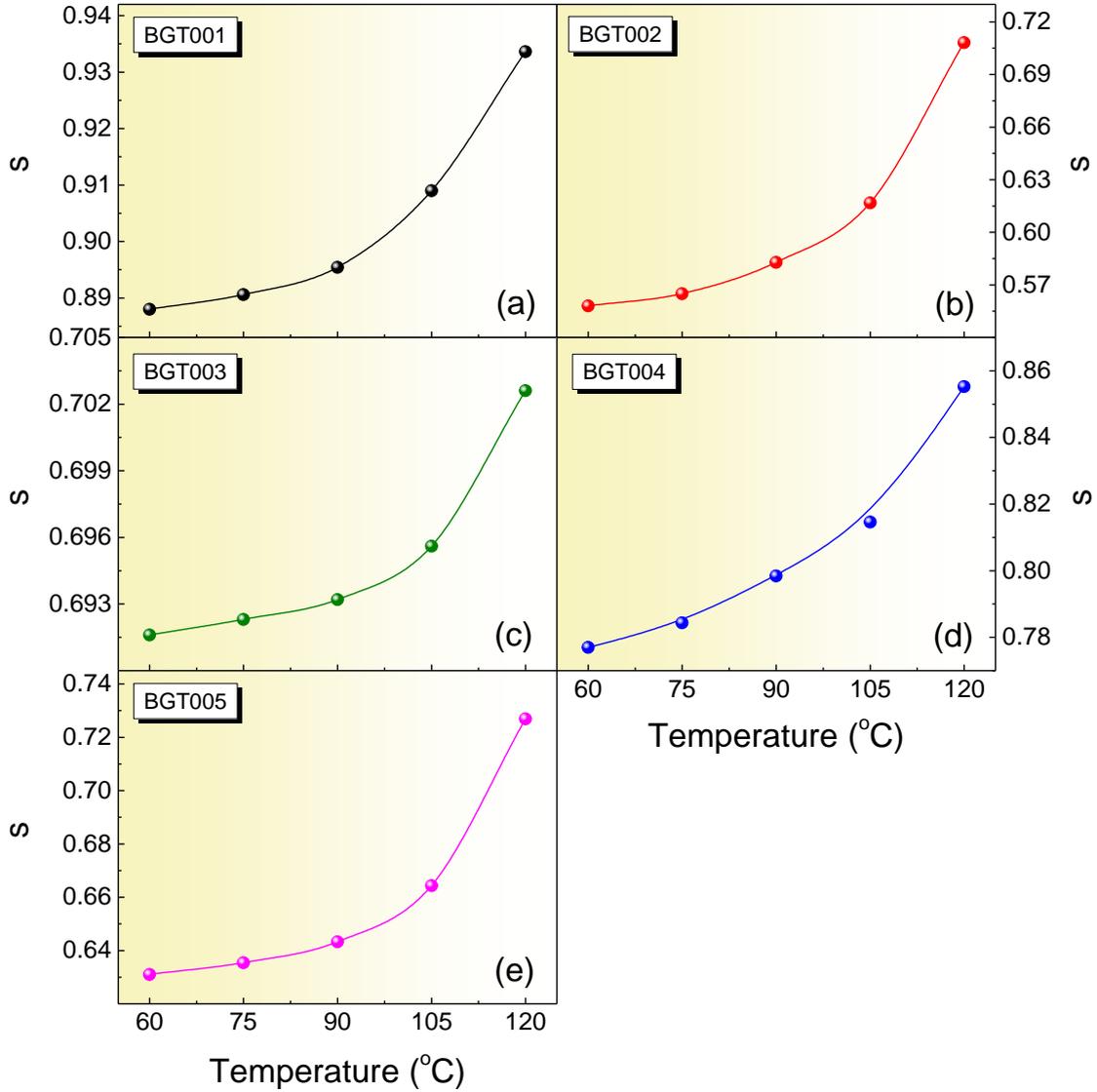

Figure 8. Temperature dependence of the frequency exponent *s*, obtained from the fitting with the Eq. (4).

In order to validate these results, additional analyses have been considered by applying the polaron hopping (PH) model [78,79], where the electrical conductivity is based on thermally activated charge carriers. According to this model, the conductivity response obeys an Arrhenius's relation, expressed by the Eq. 5 [78], where $\sigma_0$ represents the pre-exponential factor, *T* is the absolute temperature (expressed in Kelvin, K), $k_B$ is the Boltzmann constant, $E_{DC}$ is the activation energy for the DC conduction, and *n* represents an exponent factor, which determines the adiabatic (non-adiabatic) condition for the charge transport.

$$\sigma_{DC} = \frac{\sigma_0}{T^n} e^{-\frac{E_{DC}}{k_B T}} \qquad (5)$$

For adiabatic regime charges transport, as in the case of the SPH mechanism [80,81], the Eq. 5 applies when $n = 1$. Therefore, the temperature dependence of the DC conductivity values ($\sigma_{DC}$), extracted from the fitting of the experimental data of Fig. 7 with the Eq. 5, has been plotted and results are shown in Fig. 9 for all the studied compositions.

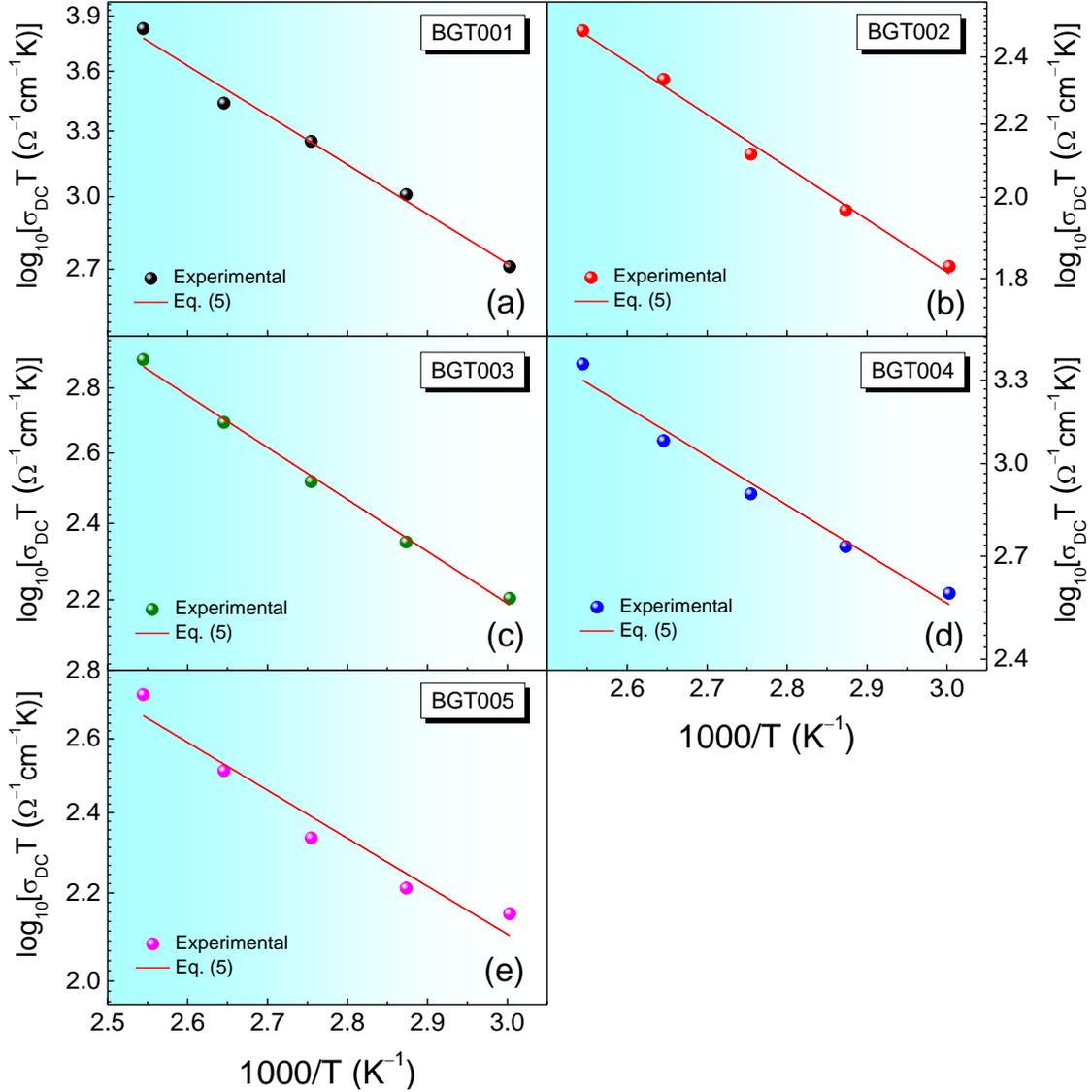

Figure 9. Temperature dependence of the of the DC conductivity ($\sigma_{DC}$), extracted from the fitting of the experimental data in Fig. 7 with the Eq. (4). Solid lines represent the fitting with the polaron hopping (PH) model relation, according to the Eq. (5).

For all the cases, the fitting process with Eq. 5 (represented by solid lines) is also shown in Fig. 9, and results clearly show that the converse variation of $\sigma_{DC}$ with temperature successfully obeys the linear behavior followed by the Arrhenius's law for the polaron hopping (PH) model, from which the activation energies for the DC conduction process have been estimated to be in the 0.22–0.31 eV interval, as reported in Table 1. As observed, the obtained values for $E_{DC}$ are relatively close to those previously obtained from the dielectric response (by using the Davidson-Cole relaxation distribution function), which indicates that the conduction processes are dominated by the same hopping mechanism of charge carriers for the electrical conduction and, therefore, corroborate that the real nature for the observed frequency dielectric dispersions is indeed predominantly governed by small polaron hopping conduction mechanism [22,82–85]. The obtained values for the activation energies from the DC conductivity analysis also suggest the role of the polaron formation and energy in the studied system and are in agreement with the reported ones for other perovskite-type oxides [13,86–89]. The slight increase in the $E_{DC}$ values with respect to the $E_a$ ones, are related to the required energy for the polaron release. In fact, according to Iguchi *et al.* [15], the shift in the temperature corresponding to the maximum dielectric permittivity ($T_m$) with the increase of frequency (ω) in the dielectric relaxation process obeys an Arrhenius' type relation, according to the Eq. (6), where $\omega_0$ is a pre-exponential factor, $k_B$ is the Boltzmann's constant, $E_a$ represents the activation energy for the polaron hopping process and $E_D$ is commonly known as the disorder energy [15]. However, since in practice $E_D$ is in the order (and even lower) of the experimental error, in most of the cases it can be neglected [82,90].

$$\omega = \omega_0 e^{-\frac{(E_a + \frac{E_D}{2})}{k_B T_m}} \tag{6}$$

Thus, taken into account that the total activation energy for the DC conduction process ($E_{DC}$) can be also affected by the contribution of the polaron hopping process ($E_a$) as well as the activation energy required for the polaron creation ($E_0$), also referred as the necessary energy for polaron to pass from the bound to the free state (that is, to create a free hopping polaron) [82,83], the contribution of the activation energy for the DC conduction process ($E_{DC}$) from Eq. (5) can be also expressed by the Eq. (7) [15].

$$E_{DC} = E_a + \frac{E_D}{2} + \frac{E_0}{2} \tag{7}$$

Therefore, by using the Eq. (7), the energy required to create a free hopping polaron ($E_0$) has been estimated and the obtained values are reported in Table 1, which reveal to be in agreement with those reported for some perovskites and other polaron-mediated conduction mechanism systems [83–85]. It is worth pointing out that the activation energy for the polaron release also shows the same trend (a decrease with the increase of the Gd concentration) to that observed for $E_a$ and $E_{DC}$, thus confirming the influence of the doping content on the defects' formation mechanism in the donor-doped $BaTiO_3$ system.

## 4. Conclusions

The conduction mechanisms of $Ba_{1-x}Gd_xTiO_3$ ceramic samples were carefully investigated using both dielectric dispersion and conductivity formalisms. The Davidson-Cole model, used to investigate the dielectric processes in the low temperature region ($T<T_C$) revealed the presence of a relaxation effect in these materials promoted by the electron hopping mechanism, yielding activation energy values in the 0.21–0.29 eV range. This result suggested that the observed relaxation processes are dominated by thermally-activated polaron mechanism due to the reduction of $Ti^{4+}$ to $Ti^{3+}$ ions for the studied BGT00$x$ system. On the other hand, through the Jonscher's universal power-law formalism, the conductivity analysis confirmed the decisive role of the small polaron hopping (SPH) charge transport for the conduction mechanism, which indeed has been associated with a local distortion in the $Ti^{4+}$–O–$Ti^{3+}$ bonds within the $BO_6$ networks as the polaron propagates through the crystal.

**Authors' contribution**

**T.H.T. Rosa:** Data curation, Formal analysis, Investigation, Visualization. **M.A. Oliveira:** Data curation, Formal analysis, Investigation. **Y. Mendez-González:** Data curation, Formal analysis, Investigation, Methodology, Writing – original draft. **F. Guerrero:** Data curation, Formal analysis, Investigation, Methodology, Writing – original draft. **R. Guo:** Data curation, Formal analysis, Investigation, Methodology. **A.S. Bhalla:** Data curation, Formal analysis, Investigation, Methodology. **J.D.S. Guerra:** Conceptualization, Data curation, Formal analysis, Investigation, Methodology, Funding acquisition, Writing - original draft, Writing - review & editing.

**Acknowledgments**

The authors thank the National Council of Scientific and Technological Development (CNPq) grants 309494/2022-2 and 408662/2023-9, Minas Gerais Research Foundation (FAPEMIG) grants PPM-00661-16 and APQ-02875-18, and Coordenação de Aperfeiçoamento de Pessoal de Nível Superior - Brasil (CAPES) - Finance Code 001, from Brazil, and NSF (01002380) and ONR (N000141613096) from the USA for the financial.